\documentclass[smallabstract,smallcaptions]{dccpaper}

\usepackage{epsfig}
\usepackage{color}
\usepackage{url}
\usepackage{amsthm,amsmath,amssymb}
\usepackage{mathrsfs}
\usepackage{graphics}
\usepackage{array}
\usepackage{multicol}
\usepackage{multirow}
\usepackage{booktabs}
\newlength{\figurewidth}
\newlength{\smallfigurewidth}
\usepackage{algorithm,algorithmic}
\usepackage[numbers]{natbib}

\begin{document}

\title{Takin-VC: Expressive Zero-Shot Voice Conversion via Adaptive Hybrid Content Encoding and Enhanced Timbre Modeling}


\author{
  Yuguang Yang$^{1*}$, Yu Pan$^{2*}$, Jixun Yao$^{3*}$, Xiang Zhang$^{1*}$, Jianhao Ye$^{1}$, \\ 
Hongbin Zhou$^{1}$, Lei Xie$^{3\dagger}$, Lei Ma$^{4\dagger}$, Jianjun Zhao$^{2}$\\ [0.5em]
  $^{1}$Ximalaya Inc., China \\
  $^{2}$Kyushu University, Japan \\
  $^{3}$Northwestern Polytechnical University, China \\
  $^{4}$The University of Tokyo, Japan \\
}



\maketitle
\thispagestyle{empty}

\begin{abstract}
Expressive zero-shot voice conversion (VC) is a critical and challenging task that aims to transform the source timbre into an arbitrary unseen speaker while preserving the original content and expressive qualities.
Despite recent progress in zero-shot VC, there remains considerable potential for improvements in speaker similarity and speech naturalness. Moreover, existing zero-shot VC systems struggle to fully reproduce paralinguistic information in highly expressive speech, such as breathing, crying, and emotional nuances, limiting their practical applicability.
To address these issues, we propose Takin-VC, a novel expressive zero-shot VC framework via adaptive hybrid content encoding and memory-augmented context-aware timbre modeling. 
Specifically, we introduce an innovative hybrid content encoder that incorporates an adaptive fusion module, capable of effectively integrating quantized features of the pre-trained WavLM and HybridFormer in an implicit manner, so as to extract precise linguistic features while enriching paralinguistic elements.
For timbre modeling, we propose advanced memory-augmented and context-aware modules to generate high-quality target timbre features and fused representations that seamlessly align source content with target timbre.
To enhance real-time performance, we advocate a conditional flow matching model to reconstruct the Mel-spectrogram of the source speech.
Experimental results show that our Takin-VC consistently surpasses state-of-the-art VC systems, achieving notable improvements in terms of speech naturalness, speech expressiveness, and speaker similarity, while offering enhanced inference speed. 
\end{abstract}


\section{Introduction}
\label{sec:intro}

{
\let\thefootnote\relax
\footnote{$*$ denotes equal contributing.}
\footnote{$\dagger$ denotes the corresponding author.}
}

Zero-shot voice conversion (VC) aims to modify the timbre of a source speech to match that of a previously unseen speaker, while maintaining the original phonetic content, has found broad applications in various practical domains \cite{gan2022iqdubbing,tomashenko2022voiceprivacy,liu2021any}.

The advancement of deep learning techniques has significantly propelled the development of zero-shot VC, with numerous methods \citep{li2023dvqvc,hussain2023ace,choi2023diff,anastassiou2024voiceshop,luo2024posterior} exhibiting impressive results in converting natural and realistic speech. 
The key idea behind is factorizing speech into distinct elements, such as content and timbre elements, and leveraging the source speech content alongside the target timbre to synthesize the desired output.
In this paradigm, the quality of content and timbre features, as well as the quality of their disentanglement, critically influences performance. Consequently, various studies have focused on developing advanced modules \citep{wu2020vqvc,wu2020one,tang2022avqvc,wang2021vqmivc,yang2022speech,huang2023limi} and information disentanglement approaches \citep{zhao2022disentangling,tang2022avqvc,dang2022training,yao2024promptvc} to enhance zero-shot VC.
However, achieving high-quality decoupling of utterances into distinct components remains challenging \citep{pan2023msac,pan2024gemo,pan2024gmp,yao2024musa}, with existing systems still exhibiting subpar performance for unseen speakers.
Two main issues are prevalent.
First, current methods cannot fully mitigate the impact of source timbre during source content extraction, a problem referred to as "timbre leakage". 
Second, these approaches often use pre-trained speaker-verification (SV) models to capture target timbre features as globally time-invariant representations. 
Nonetheless, such SV embeddings cannot ensure robust timbre modeling and vary with linguistic content \citep{jiang2024mega,pan2024ctefm} which may diminish their effectiveness.

Recently, the progressions in large-scale speech language models \citep{wang2023lm,borsos2023audiolm} have tried to tackle this issue by leveraging robust in-context learning capabilities for converting target speech from concise utterances as prompts. Nevertheless, these methods suffer from stability issues and error accumulation due to their auto-regressive nature, which can gradually degrade conversion quality.
Moreover, current state-of-the-art (SOTA) zero-shot VC systems still struggle to simultaneously transfer the paralinguistic characteristics in highly expressive speech, such as crying, breathing, and emotional nuances, thus limiting their effectiveness and practical applicability.

In this paper, we propose \textit{Takin-VC}, a novel expressive zero-shot VC framework that delivers advanced modeling of content, timbre, and speech quality in a zero-shot fashion.
To be specific, we introduce an adaptive fusion-based hybrid content encoder that seamlessly combines the strengths of phonetic posterior-grams (PPGs) and self-supervised learning (SSL)-based representations derived from pre-trained HybridFormer \citep{yang2023hybridformer} and WavLM \citep{chen2022wavlm}.
This integration enables the precise extraction of linguistic content while simultaneously enriching paralinguistic elements. 
For timbre modeling, we first advocate a memory-augmented module capable of generating high-quality conditional target timbre inputs for our conditional flow matching (CFM) model.
To further enhance speaker similarity, a context-aware timbre modeling module based on an efficient cross-attention (CA) mechanism is presented. This module effectively aligns and fuse the extracted source content and target timbre features, rather than solely using the source linguistic content as the conditional input for CFM.
Conditioned on these features, the predicted outputs of the CFM model are ultimately fed into a pre-trained vocoder \citep{lee2022bigvgan} to synthesize the target speech.

Experiments conducted on both large-scale 500k-hour multilingual (Mandarin and English) and small-scale LibriTTS \cite{zen2019libritts} datasets demonstrate that Takin-VC consistently outperforms several SOTA zero-shot VC methods in speech naturalness, expressiveness, speaker similarity, and real-time performance. 
Notably, Takin-VC achieves significant improvements in both subjective and objective metrics compared to all baseline systems, further validating its effectiveness and robustness.
For more detailed speech samples, please visit our \textbf{demo page} \footnote{\url{https://anonymous.4open.science/w/takin-vc-0CD8/}}.
In summary, the primary contributions of this work are as follows:
\begin{itemize}
\item We present Takin-VC, a novel expressive zero-shot VC framework.
To the best of our knowledge, this is the first approach capable of simultaneously transforming the source timbre to arbitrary unseen speakers while effectively maintaining the paralinguistic characteristics of highly expressive speech.
\item We introduce an adaptive hybrid content encoder that employs an adaptive feature fusion module to implicitly integrate PPGs and quantized SSL features in a learnable manner, thereby capturing precise linguistic elements with enriched paralinguistic characteristics.
\item We propose memory-augmented and content-aware modules to enhance timbre modeling. The former aims to extract high-quality target timbre conditions, while the latter focuses on generating fused features that align and leverage target timbre embeddings with source content for the conditional flow matching model.
\end{itemize}

\section{Background}
\label{sec:Background}

\subsection{Zero-shot Voice Conversion}

Recent progressions in deep learning techniques, such as SSL-based speech models \cite{hsu2021hubert,chen2022wavlm,baevski2020wav2vec} and diffusion models \cite{ho2020denoising,lu2022dpm}, have greatly advance zero-shot VC. 
SEF-VC \cite{li2024sef} utilizes a CA mechanism to extract timbre features and reconstruct waveforms from HuBERT \cite{hsu2021hubert} tokens, while \cite{choi2023diff} proposes a diffusion-based hierarchical VC method using XLS-R \cite{babu2021xls} for content extraction and dual diffusion models for generating pitch and Mel-spectrograms.
Despite these innovations, SSL-based zero-shot VC methods \cite{dang2022training,hussain2023ace,li2023dvqvc} are likely to encounter the timbre leakage challenge, as SSL features do not explicitly disentangle timbre features. Likewise, diffusion-based approaches \cite{popov2021diffvc,choi2024dddm} suffer from suboptimal real-time performance.
Another emerging paradigm \cite{zhang2023speak,wang2023lm,baadeneural} involves decoupling speech into semantic and acoustic tokens using neural codecs \citep{defossez2022high,yang2023hifi,pan2024promptcodec} and SSL-based models, subsequently using language models to generate converted speech. 
While these approaches mark impressive results, current SOTA VC methods still have considerable room for improvement in achieving better speaker similarity and naturalness. Besides, they continue to face difficulties in faithfully and simultaneously reproducing the paralinguistic characteristics of highly expressive speech.

\subsection{Flow Matching-based Generative Models}
Flow matching-based generative models \cite{lipman2022flow,tong2023improving,tong2023simulation} have recently emerged as a powerful solution for generative tasks. By estimating vector fields to approximate the transport path from noise to the target distribution, these models employ neural ordinary differential equations (ODEs) to learn optimal transport trajectories. Compared to diffusion-based methods \cite{bartosh2023neural,zhou2023denoising}, flow matching offers improved training stability and real-time performance by enabling direct noise-to-sample mapping while significantly reducing sampling steps.
In the speech processing domain, flow matching-based systems are emerging as a promising paradigm. SpeechFlow \cite{liu2023generative} uses a pre-trained flow matching model with masked conditions on large-scale untranscribed speech data, facilitating speech enhancement and separation tasks. P-Flow \cite{kim2024p} adopts speech prompts for speaker adaptation, integrating a speech-prompted text encoder and a flow matching decoder to enable high-quality and real-time speech synthesis.
Despite these advancements, the application of flow matching in zero-shot VC remains nascent, underscoring the need for developing a stable and efficient flow matching-based zero-shot VC framework.

\section{METHODs}
\label{sec:METHODOLOGY}

\subsection{Overivew}

\begin{figure}[htbp]
    \centering
    \includegraphics[width=16cm]{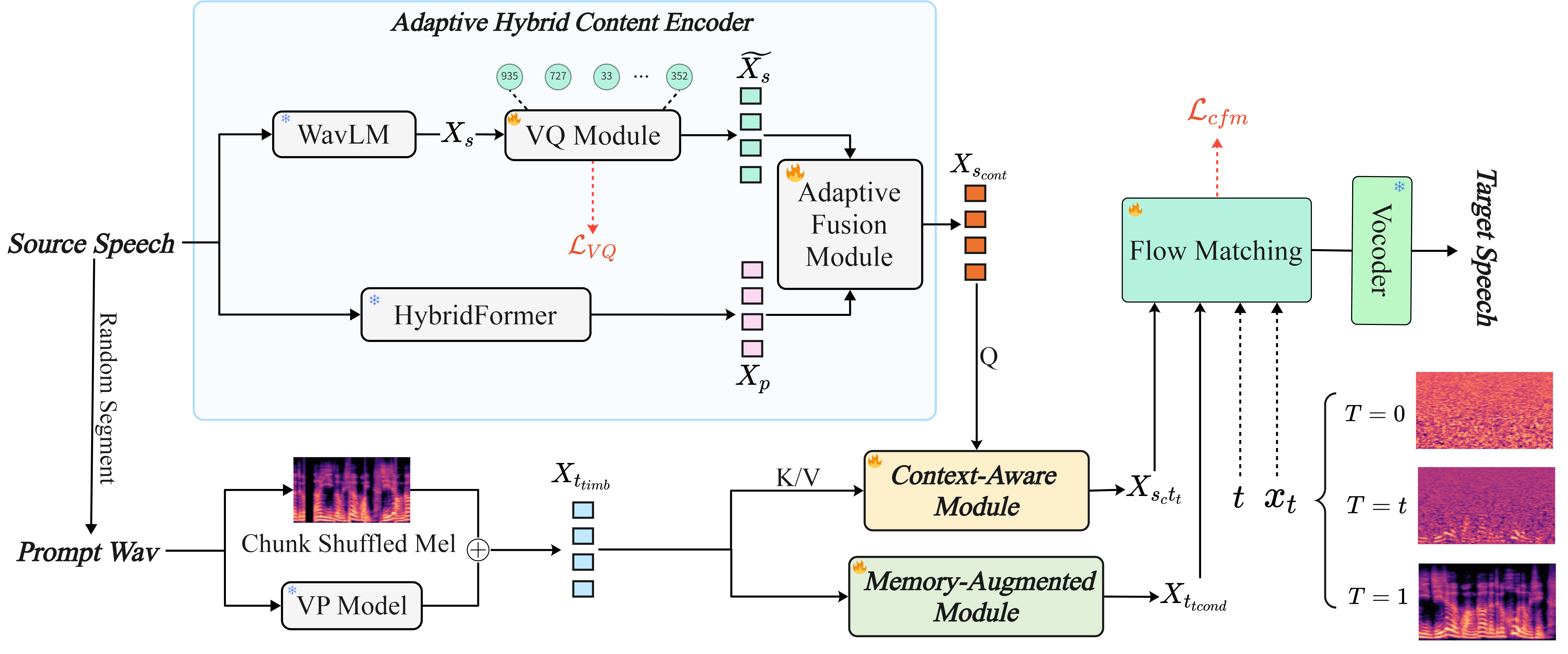}
    \caption{Overview of Takin VC.}
    \label{fig:main}
\end{figure}

As shown in Fig. \ref{fig:main}, our Takin-VC system primarily comprises three key components: an adaptive hybrid content encoder, a memory-augmented context-aware timbre modeling approach, and a conditional flow matching-based decoder.

In detail, the objective of the adaptive hybrid content encoder is to precisely capture linguistic characteristics enriched with paralinguistic elements, denoted as $X_{s_{cont}}$. To achieve this, an adaptive feature fusion module on top of the hybrid content encoder is presented to effectively leverage the complementary strengths of PPG and quantized SSL representations in a learnable fashion.
Regarding timbre modeling, we first propose a memory-augmented module that incorporates a stack of convolution, activation, and self-attention layers to extract high-quality target timbre conditions $X_{t_{tcond}}$ for the CFM model.
To further improve timbre modeling capabilities, a cross-attention-based context-aware module is presented to generate fused representations $X_{s_{ct_t}}$ that effectively integrate $X_{s_{cont}}$ with target timbre.
Finally, to enable stable training and accelerate the reference speed, we design a CFM model that consists of multiple UNet \cite{ronneberger2015u} blocks to reconstruct the source Mel-spectrograms conditioned on $X_{s_ct_t}$ and $X_{t_{tcond}}$, followed by a pretrained Bigvgan vocoder to synthesize the desired target speech.

\subsection{Adaptive Hybrid Content Encoder}
\label{sec:content-encoder}

Current mainstream zero-shot VC systems typically use pretrained automatic speech recognition (ASR) \cite{gulati2020conformer,yang2022lmec,kim2022squeezeformer} or SSL-based speech models to capture linguistic content from the original waveform. 
However, they both have inherent limitations: ASR-derived PPGs lack sufficient paralinguistic elements, whereas SSL-based models do not explicitly disentangle timbre information.
To address these flaws, we propose an adaptive fusion-based hybrid content encoder within the Takin-VC framework, integrating the merits of both approaches.

Formally, given an input source speech $X$, our adaptive hybrid content encoder separately encodes its corresponding PPG and SSL features, denoted as $X_{p}$ and $X_{s}$, using pre-trained HybridFormer and WavLM, respectively. 
To alleviate potential timbre leakage, a residual vector quantization (RVQ) based quantizer of EnCodec \cite{defossez2022high} is applied to discretize $X_{s}$, resulting in $\tilde{X_{s}}$.
Additionally, we introduce a gradient-driven adaptive feature fusion module to further reduce timbre leakage and effectively integrate the complementary benefits of PPG and SSL features. 
Unlike conventional element-wise addition for feature fusion, the proposed strategy first processes the quantized WavLM features through a multi-layer projection module comprising a one-dimensional convolutional (Conv1d) layer followed by a LeakyReLU activation function, with the negative slope empirically set to 0.2. 
Temporal interpolation is then applied to ensure dimensional alignment with the PPG features, and the resulting WavLM representations are employed as coefficients for element-wise multiplication with the PPGs:
\begin{equation} 
X_{s_{cont}} = \text{LeakyReLU}(\text{Conv1d}(\tilde{X_{s}})) \cdot X_{p} 
\label{equ:111}
\end{equation}
where Conv1d denotes the 1D convolutional layer.
By this means, as gradients propagate back to Formula \ref{equ:111} during training, the limited representation of paralinguistic nuances within the PPG features results in larger gradient magnitudes for these elements. Since the PPGs are fixed before training, the gradients primarily affect the adaptive fusion module associated with the quantized WavLM features. As a consequence, this gradient-driven adjustment dynamically optimizes the weights of the quantized WavLM features in an implicit way, thereby amplifying the representation of paralinguistic elements in the combined feature space, improving overall content modeling capabilities, and significantly reducing the risk of voiceprint leakage.

\subsection{Enhanced Timbre Modeling}
\subsubsection{Memory-augmented Timbre Modeling}

\begin{figure}[htbp]
    \centering
    \includegraphics[width=9cm]{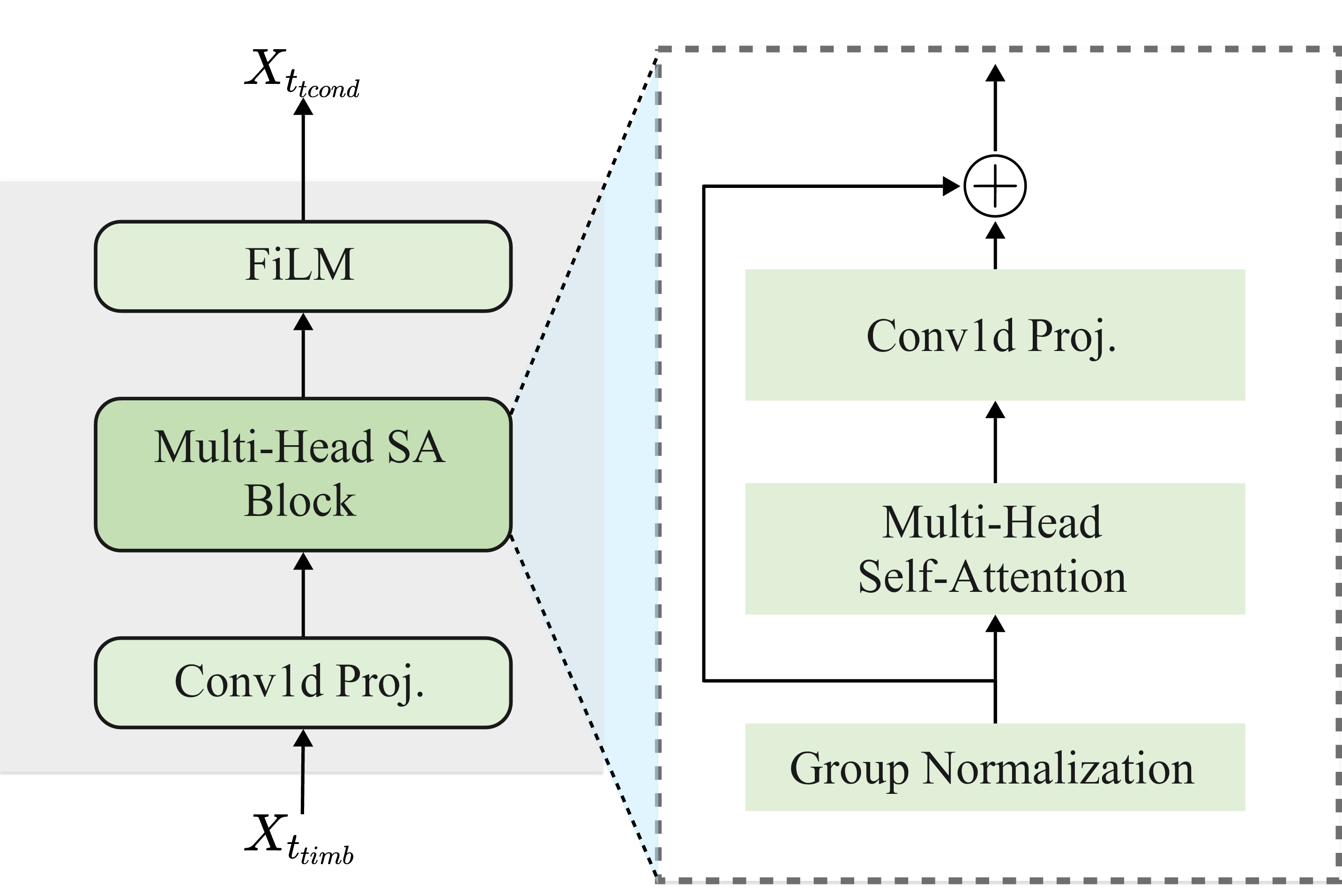}
    \caption{Schematic of the memory-augmented module.}
    \label{fig:memory-augmented-timbre-modeling}
\end{figure}

\noindent
To capture high-quality target timbre conditions for the CFM model, we propose an efficient memory-augmented module that adaptively integrates the shuffled Mel-spectrogram and VP features of the reference speech, as outlined in Fig. \ref{fig:memory-augmented-timbre-modeling}.

Detailed, we extract the Mel-spectrograms from randomly segmented reference waveforms originating from the same speaker as the source speech. The individual frames of these Mel-spectrograms are then shuffled to preserve essential timbre characteristics while minimizing the influence of the source speech content.
Subsequently, a lightweight pre-trained SV model\footnote{\url{https://modelscope.cn/models/iic/speech_campplus_sv_zh_en_16k-common_advanced}} is utilized to extract timbre embeddings from the reference speech. These embeddings are then concatenated with the shuffled Mel-spectrograms, resulting in the target timbre representations, referred to as $X_{t_{timb}}$.
To refine these concatenated features, our proposed memory-augmented module that begins by employing a Conv1d layer to project the captured features and then incorporates four SA blocks, each comprising a group normalization layer, multi-head SA mechanism, a Conv1d layer, and a shortcut connection operation. 
The resulting features are then subjected to a temporal averaging operation, followed by the application of a FiLM layer \cite{perez2018film} to perform affine feature-wise transformation, producing the conditional target timbre inputs $X_{t_{tcond}}$.

\subsubsection{Context-aware Timbre Modeling}

Speaker timbre features have long been viewed as global and time-invariant representations \citep{lin2021s2vc,li2024sef,pan2024ctefm}. However, recent studies \citep{jiang2024mega} have revealed a close interdependence between timbre modeling and content information.
Hence, drawing inspiration from this insight, we propose an innovative context-aware timbre modeling approach based on advanced cross-attention mechanism.

\begin{figure}[htbp]
    \centering
    \includegraphics[width=9cm]{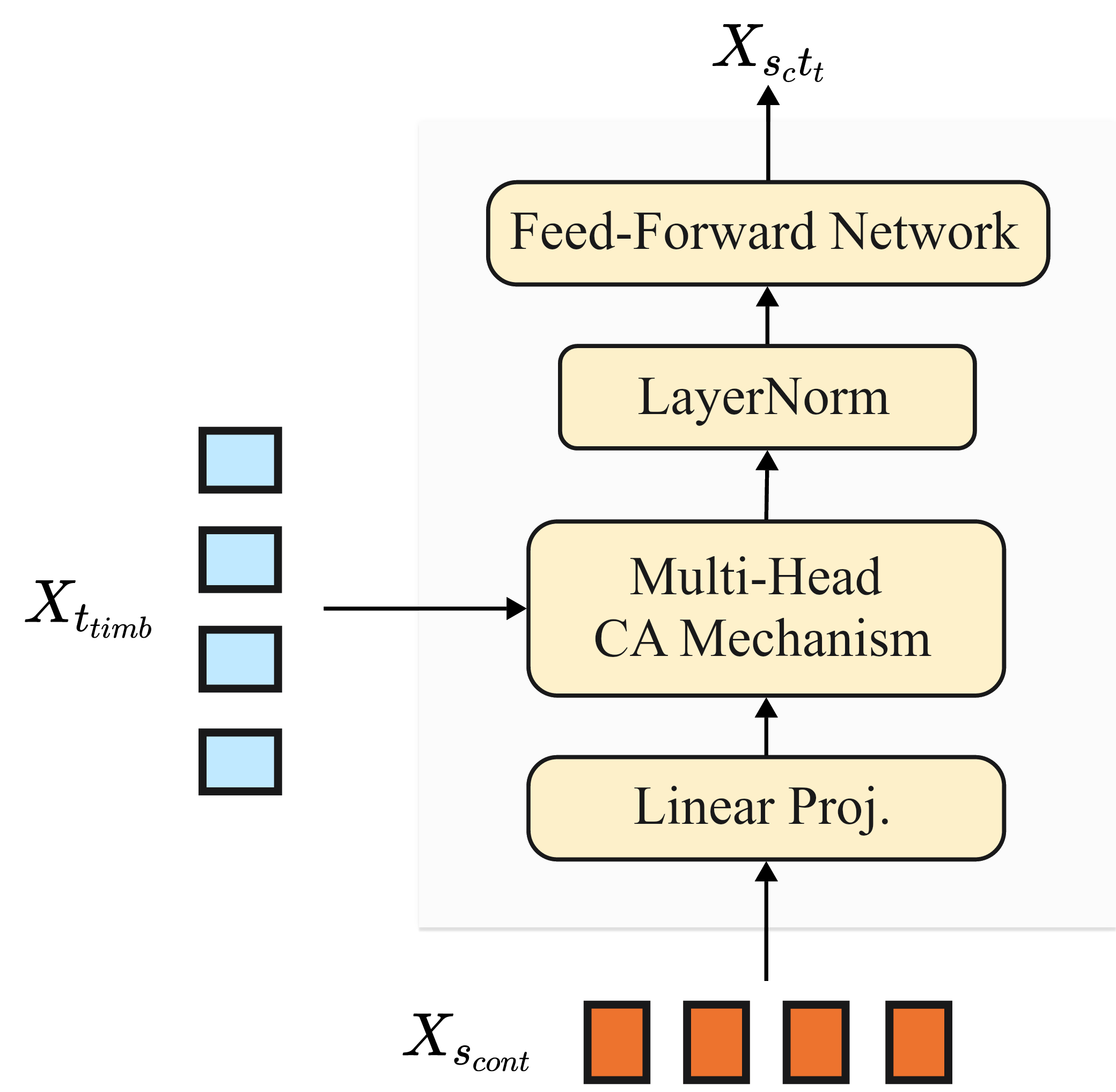}
    \caption{Schematic of the context-aware module.}
    \label{fig:context-aware-timbre-modeling}
\end{figure}

As illustrated in Fig. \ref{fig:context-aware-timbre-modeling}, the CA-based module is designed to generate semantically aligned timbre features that harmonize the source linguistic content with the target timbre. 
Concretely, the CA-based module consists of a series of linear projection layers, multi-head cross-attention layers, layer normalization, and position feed-forward network (FFN), which can effectively facilitate the integration of $X_{s_{cont}}$ and $X_{t_{timb}}$.
The source content \(X_{s_{cont}}\) is used as the query, while the target timbre \(X_{t_{timb}}\) serves as both the key and value.
Finally, the extracted features $X_{s_ct_t}$ are interpolated to ensure dimensional compatibility with the ground truth, i.e., the source Mel-spectrogram, facilitating the subsequent training of the CFM model.

\subsection{Conditional Flow Matching-based Decoder} 

In Takin-VC, to facilitate more efficient training and faster inference, we use a CFM model with optimal-transport (OT-CFM) to approximate the distribution of source Mel-spectrograms and generate predicted outputs conditioned on $X_{s_{ctt}}$ and $X_{t_{tcond}}$, all in a simulation-free manner.

Assume that the standard distribution and the target distribution are denoted as $p_0(x)$ and $p_1(x)$, respectively. 
The OT flow $\phi: [0, 1] \times R^d \rightarrow R^d$ establishes the mapping between two density functions through the use of an ordinary differential equation (ODE):
\begin{equation}
    \label{eq:ode}
    \begin{split}
        \frac{d}{d_t} \phi_t(x)=v_t(\phi_t(x),t) \\
        \phi_0(x)\sim p_0(x)=\mathcal{N}(x;0,I), \quad \phi_1(x)&\sim p_1(x) \\
    \end{split}
\end{equation}

where $v_t$ is a learnable time-dependent vector field, and $t \in [0, 1]$. 
Since multiple flows can generate this probability path, making it challenging to determine the optimal marginal flow, we adopt a simplified formulation, as proposed in \cite{tong2023conditional}:
\begin{equation}
    \label{eq:111}
    \begin{split}
        \phi^{OT}_{t,z}(x) = \mu_t(z) + \sigma_t(z)x \quad \quad \quad \\ 
        \mu_t(z) = (1 \!-\! (1\!-\!\sigma_{min})t) z, \quad \sigma_t(z) = t \\
    \end{split}
\end{equation}

where $z$ represents the random variable, $\sigma_{min}$ is a hyper-parameter set to 0.0001. 
Therefore, the training objective of the proposed CFM model can be formulated as:

\begin{equation}
    \mathcal{L}_{takin} = \mathbb{E}_{t,p(x_0),q(x_1)}\Vert((x_1-(1-\sigma)x_0)-v_t({\phi^{OT}_{t,x_1}}(x_0)|\theta,h)\Vert^2
\end{equation}

where $\theta$ represents the parameters of the flow matching model, and \(h\) denotes the conditional set comprising \(X_{t_{tcond}}\) and \(X_{s_ct_t}\).

\subsection{Training Objective}

The training objective of the proposed Takin-VC is composed of two components, i.e., the RVQ commitment loss $\mathcal{L}_{vq}$ of the VQ module and $\mathcal{L}_{cfm}$.
\begin{equation}
    \begin{gathered}
        \mathcal{L}_{total} = \mathcal{L}_{cfm} + \lambda \mathcal{L}_{vq} \\
        \mathcal{L}_{vq}(X_{s}, \tilde{X_{s}}) = \sum_{i=1}^N \left\| X_{s_i} - \hat{X}_{s_i} \right\|_2^2
    \end{gathered}
\end{equation}

Here, \(\lambda\) is a hyper-parameter that controls the weight of \(L_{vq}\), and \(N\) represents the number of RVQ-based quantizers. In our implementation, \(\lambda\) is empirically set to 0.01, \(N\) is set to 1, and the codebook size of the RVQ-based quantizer is empirically determined to be 8200.

\section{Experimental Setup}
\label{sec:setup}

\subsection{Baseline System}
We conduct a comparative experiment of the performance in zero-shot voice conversion between our proposed Takin-VC approach and baseline systems, encompassing the following system: 1) DiffVC~\citep{popov2021diffvc}: A zero-shot VC system based on diffusion probabilistic modeling, which employs an averaged mel spectrogram aligned with phoneme to disentangle linguistic content and timbre information; 
2) NS2VC\footnote{\url{https://github.com/adelacvg/NS2VC}}: A modified voice conversion edition of NaturalSpeech2~\citep{shen2023naturalspeech2}, which employ both diffusion and codec model to achieve zero-shot VC;
3) VALLE-VC~\citep{wang2023valle}: We replace the original phoneme input with the semantic token extracted from the supervised model to make VALLE convert the timbre of source speech to the target speaker;
4) SEFVC~\citep{li2024sef}: A speaker embedding free voice conversion model, which is designed to learn and incorporate speaker timbre from reference speech.
5) StableVC~\citep{yao2024stablevc}: A style controllable zero-shot voice conversion system, which employs dual adaptive gate attention to capture timbre and style information.
6) SeedVC~\citep{liu2024seedvc}: A zero-shot voice conversion system with an external timbre shifter and diffusion transformer.

\subsection{Evaluation Metrics}
Both subjective and objective metrics are employed to evaluate the performance of our Takin-VC and baseline systems. For \textbf{subjective metrics}, we employ naturalness mean opinion score (NMOS) to evaluate the naturalness of the generated samples and similarity mean opinion scores (SMOS) to evaluate the speaker similarity. We invite 20 professional participants to listen to the generated samples and provide their subjective perception scores on a 5-point scale: '5' for excellent, '4' for good, '3' for fair, '2' for poor, and '1' for bad.
For \textbf{objective metrics}, we employ word error rate (WER), UTMOS, and speaker embedding cosine similarity (SECS) to evaluate the intelligibility, quality, and speaker similarity. 
Specifically:
1) We use a pre-trained CTC-based ASR model\footnote{\url{https://huggingface.co/facebook/hubert-large-ls960-ft}} to transcribe the generated speech and compare with ground-truth transcription;
2) We use a MOS prediction system that ranked first in the VoiceMOS Challenge 2022\footnote{\url{https://github.com/tarepan/SpeechMOS}} to estimate the speech quality of the generated samples;
3) We use the WavLM-TDCNN SV model\footnote{\url{https://github.com/microsoft/UniSpeech/tree/main/downstreams/speaker_verification}} to measure speaker similarity between generated speech and target speech. 
Furthermore, we introduce real-time factor (RTF) to evaluate the efficiency of Takin-VC. 

\subsection{Dataset}

\subsubsection{Small Scale Dataset}

We employ the LibriTTS dataset to train our system and baseline systems, which contain 585 hours of recordings from 2,456 English speakers. We follow the official data split, using all training datasets for model training and "dev-clean" for model selection. The "test-clean" dataset is used to construct the evaluation set. All samples are processed at a 16kHz sampling rate.

\subsubsection{Large Scale Dataset}

To train a robust Takin VC model, we collected a dataset of approximately 500k hours. During the data collection process, we used an internally constructed data pipeline specifically designed for audio large model tasks. This pipeline includes signal-to-noise ratio (SNR) filtering, audio spectrum filtering (filtering out 24k audio with insufficient high-frequency information and pseudo 24k audio), VAD (Voice Activity Detection), LiD+ASR (Language Identification + Automatic Speech Recognition), speaker separation and identification, punctuation prediction, and background noise filtering. Regarding the test set, to validate the effectiveness of the Takin-VC model, we collected speech data from the internet that includes 100 non-preset speakers for evaluation. These speakers represent a variety of attributes such as gender, age, language, and emotion to ensure a comprehensive evaluation of the model's performance.

\subsection{Model Configuration}

For the content encoder part, in the first stage, we used the 12-layer HybridFormer-base model trained on a large dataset of 500K hours. For the WavLM part, we used the output features of the 6th layer. In the VQ part, we adopted a single-layer 8200 codebook with a hidden dimension of 1024, trained for 1 million steps on 100K hours of data. The fusion layer, as described in Sec. \ref{sec:content-encoder}, is a simple module with several convolutional layers, an activation layer, and weighted summation. The Decoder adopts the same structure and configuration as HiFi-codec \cite{yang2023hifi}.

In the part of timbre modeling and flow matching model, both the context-aware and memory-augmented modules use a transformer block with 8 heads, 6 layers, and a hidden size of 1024, with only the form of attention being different. The main structure of CFM uses a design of 10-layer U-net plus 3 layers of ResNet block~\citep{he2016identity}, with a hidden size of 1280. A Memory Fusion Block is inserted into the 10-layer U-net to enhance the speaker similarity of the generated audio.


For the small-data experiments, we use four A800 GPUs, whereas the large-data experiments are conducted on eight A800 servers. The batch size on each GPU is set to 16 with the AdamW optimizer using 1e-4 as the learning rate. In the inference section, experiments typically took 5 to 20 steps, with the final table uniformly adopting the results of 10 steps. The Classifier-Free Guidance (CFG) coefficient ranged from 0.1 to 1.0, with 0.7 used in the table. The specific experimental results will be detailed later.

\section{Experimental Results}
\subsection{Experiments on small dataset}

\begin{table*}[ht]
\centering
\caption{Comparison results of subjective and objective metrics between Takin-VC and the baseline systems in zero-shot voice conversion. Subjective metrics are computed with 95\% confidence intervals and ``GT" refers to ground truth samples.}\label{tab:libri_comp}
\begin{tabular}{lcccccc}
\hline
         & NMOS ($\uparrow$)& SMOS ($\uparrow$) & WER ($\downarrow$) & UTMOS ($\uparrow$) & SECS ($\uparrow$) & RTF ($\downarrow$)\\ \hline
GT       & 4.17$\pm0.04$ & -    & 2.04 & 4.21  & -    \\ \hline
DiffVC   & 3.75$\pm0.05$ & 3.66$\pm0.07$ & 3.08 & 3.68  & 0.61   & 0.294 \\
NS2VC    & 3.65$\pm0.07$ & 3.51$\pm0.06$ & 2.94 & 3.64  & 0.53   & 0.347\\
VALLE-VC & 3.80$\pm0.06$ & 3.79$\pm0.04$ & 2.77 & 3.72  & 0.65   & 3.678\\
SEFVC    & 3.68$\pm0.05$ & 3.76$\pm0.06$ & 3.75 & 3.51  & 0.63   & 0.187 \\
StableVC    & 3.83$\pm0.04$ & 3.88$\pm0.06$ & 2.77 & 3.92  & 0.66   & 0.267 \\
SeedVC    & 3.87$\pm0.05$ & 3.74$\pm0.06$ & 2.51 & 3.81  & 0.68   & 0.341 \\
\hline
Takin-VC & \textbf{3.98$\pm0.04$} & \textbf{4.11$\pm0.05$} & \textbf{2.35} & \textbf{4.08}  & \textbf{0.71} & \textbf{0.154}\\ \hline
\end{tabular}
\end{table*}

We first evaluate the performance of our proposed Takin-VC using subjective metrics. These metrics capture human perception of the enhanced speech's naturalness, intelligibility, and speaker similarity. As shown in Table~\ref{tab:libri_comp}, we can find that 1) our proposed system achieves the highest NMOS of 3.98, which is significantly higher than baseline systems; 2) the speaker similarity of our proposed system also outperforms all baseline systems. These results demonstrate that Takin-VC can achieve superior performance than the baseline system in the perceived aspect.

Furthermore, we evaluate the performance using objective metrics. The WER of our proposed system is 2.35, only slightly higher than the ground truth samples, indicating that the samples generated by Takin-VC exhibit better intelligibility. Moreover, Takin-VC achieves a UTMOS of 4.08 and an SECS of 0.71, demonstrating superior quality and similarity performance. Overall, the objective results of our proposed Takin-VC outperform all baseline systems and further corroborate the subjective findings. For inference efficiency, Takin-VC achieves the lowest RTF over all baseline systems, demonstrates superior real-time performance.

\subsection{Experiments on large dataset}

\begin{table*}[ht]
\centering
\caption{Detailed results of Takin-VC on different conversion scenarios. ``F'' and ``M'' represent the female and male, respectively.}\label{tab:large}
\begin{tabular}{lccccc}
\hline
    & NMOS ($\uparrow$)& SMOS ($\uparrow$)& WER  ($\downarrow$)& UTMOS ($\uparrow$)& SECS ($\uparrow$)\\ \hline
GT  & 4.21$\pm0.05$ & - & 2.11 & 4.18  & -    \\ \hline
F2F & 4.16$\pm0.04$ & 4.18$\pm0.03$ & 2.11 & 4.11  & 0.74 \\
F2M & 4.14$\pm0.05$ & 4.09$\pm0.05$ & 2.24 & 4.13  & 0.71 \\
M2M & 4.12$\pm0.04$ & 4.11$\pm0.04$ & 2.20 & 4.20  & 0.73 \\
M2F & 4.13$\pm0.05$ & 4.04$\pm0.06$ & 2.31 & 4.09  & 0.70 \\ \hline
\end{tabular}
\end{table*}

We employ the large scale dataset to train our proposed Takin-VC and investigate the performance in different conversion scenarios across different gender. As shown in Table \ref{tab:large}, we divide the experiments into four groups: female to female (F2F), female to male (F2M), male to male (M2M), and male to female (M2F) to investigate performance differences. The results show that all metrics outperform Takin-VC trained on a smaller dataset, demonstrating that our proposed approach scales effectively. Additionally, the conversion results for same-gender conversions are slightly better than cross-gender conversions in both SMOS and SECS, while other metrics remain similar across all four group settings.

\begin{figure}[h]
    \centering
    \includegraphics[width=16cm]{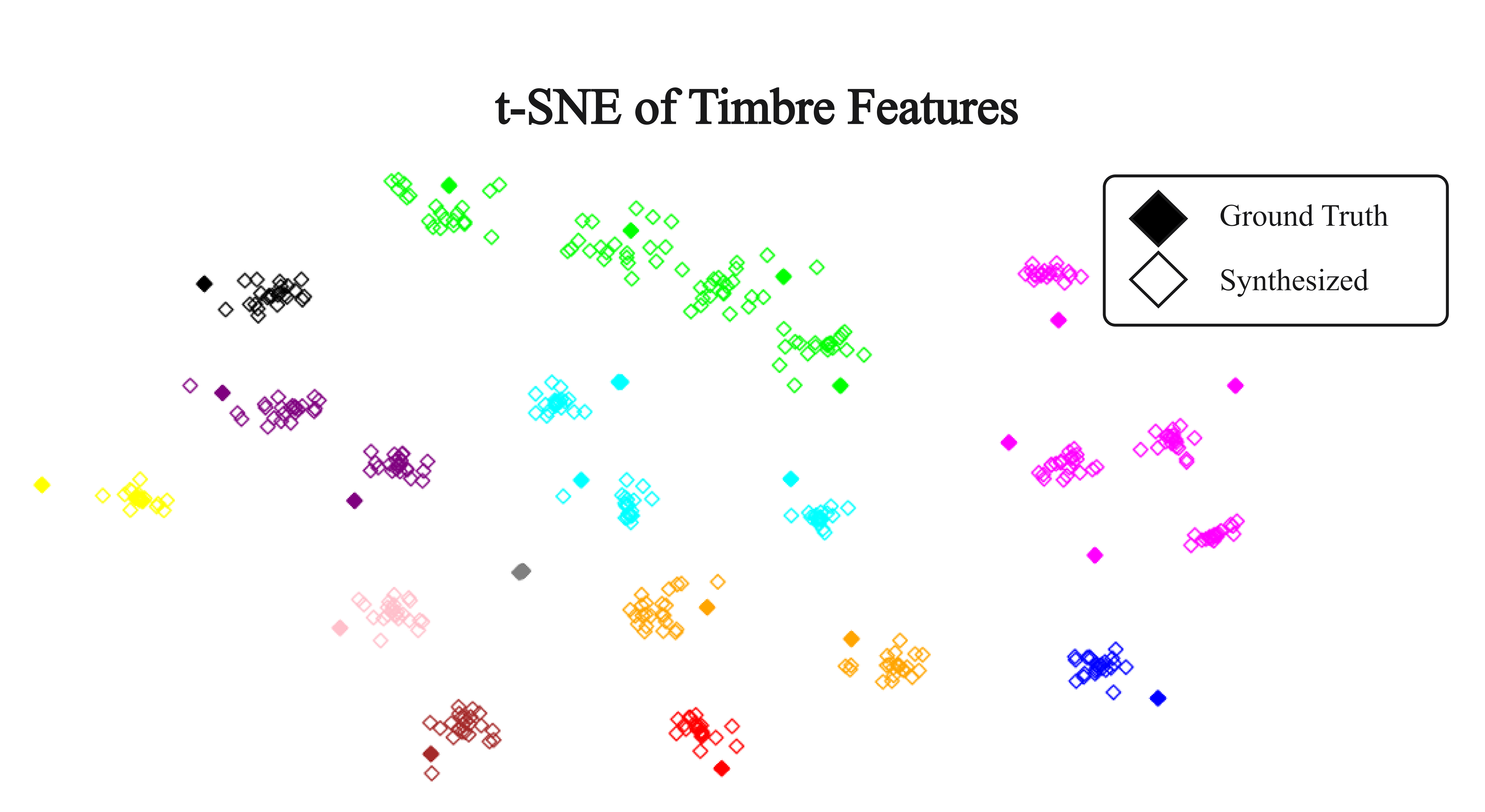}
    \caption{The t-SNE result of speaker similarity between ground truth samples and converted speech.}
    \label{fig:tsne}
\end{figure}

To further investigate the speaker similarity performance of our Takin-VC, we use the t-SNE method~\citep{van2008visualizing} to visualize the speaker embeddings of 13 speakers, comparing the ground truth samples with the converted samples generated by Takin-VC. 
As shown in Figure \ref{fig:tsne}, the embeddings of real and converted speech from the same speaker are closely clustered. This demonstrates that the speech generated by Takin-VC closely matches real human speech in both quality and speaker similarity.

\subsection{Ablation Study}

\begin{table}[h]
\centering
\caption{The ablation results for linguistic content extraction modules. ``w/o ppg'' and ``w/o SSL'' represent removing the HybridFormer or WavLM branch in our proposed hybrid content encoder, respectively.}\label{tab:ablation_content}
\begin{tabular}{lccccc}
\hline
         & NMOS & SMOS  & WER & UTMOS  & SECS  \\ \hline
Takin-VC & 3.98$\pm0.04$ & 4.11$\pm0.05$ & 2.35 & 4.08  & 0.71 \\
 \quad w/o ppg & 3.74$\pm0.04$ & 3.07$\pm0.04$ & 2.79 & 3.91  & 0.45 \\
 \quad w/o SSL & 3.63$\pm0.04$ & 3.81$\pm0.04$ & 2.64 & 3.84  & 0.67 \\
\hline
\end{tabular}
\end{table}

We conduct two ablation experiments to evaluate the effectiveness of each proposed component in linguistic content extraction and timbre modeling. 
As shown in Table \ref{tab:ablation_content}, SMOS results are significantly degraded, suggesting that only using the SSL model to extract linguistic content will result in timbre leakage. When we remove the SSL model in the hybrid content encoder and only use HybridFormer to extract linguistic content, we can find that NMOS and WER results degrade. This suggests that the conventional ASR encoder is less capable of disentangling linguistic content from the necessary paralinguistic information, underscoring the importance and effectiveness of our hybrid encoder in extracting linguistic content.

\begin{table}[h]
\centering
\caption{The ablation results for timbre-related modules. ``w/o con'' represents removing content-aware timbre modeling and only employing voice print to extract timbre information. ``w/o vp'' represents removing the voice print, and ``w/o mem'' means removing the memory-augmented timbre modeling module.}\label{tab:ablation_timbre}
\begin{tabular}{lccccc}
\hline
         & NMOS & SMOS  & WER & UTMOS  & SECS  \\ \hline
Takin-VC & 3.98$\pm0.04$ & 4.11$\pm0.05$ & 2.35 & 4.08  & 0.71 \\
 \quad w/o con & 3.77$\pm0.04$ & 3.61$\pm0.04$ & 3.01 & 3.85  & 0.58 \\
 \quad w/o vp & 3.94$\pm0.05$ & 3.89$\pm0.04$ & 2.51 & 3.98  & 0.61 \\
 \quad w/o mem & 3.92$\pm0.04$ & 3.75$\pm0.05$ & 2.44 & 4.01  & 0.52 \\
\hline
\end{tabular}
\end{table}

Additionally, we conduct an ablation study for timbre-related modules, results are shown in Table~\ref{tab:ablation_timbre}. We find significant degradation across all metrics when removing context-aware timbre modeling. It suggests that the system can not capture timbre information as well without the module, resulting in poor generation results.
We observe a notable decline in speaker similarity when the voice print is removed from the attention module. We believe the voice print introduces a stronger timbre bias, which helps the attention module focus on capturing timbre information. Furthermore, when we remove the memory-augmented timbre modeling module, SMOS and SECS scores show significant degradation compared to the original Takin-VC, demonstrating the critical role of the memory module in improving timbre modeling. These ablation results demonstrate the effectiveness of each component proposed in our Takin-VC.

\section{Conclusion}
\label{sec:CONCLUSIONs}

In this study, we introduce Takin-VC, an effective framework for expressive zero-shot VC. Leveraging an adaptive fusion-based hybrid content encoder, Takin-VC integrates the complementary strengths of PPGs and quantized WavLM features in a learnable manner, thereby enhancing the naturalness and expressiveness of the converted speech.
To improve speaker similarity, we propose an advanced memory-augmented module capable of extracting fine-grained conditional target timbre features. Additionally, we design a context-aware timbre modeling module to capture fused representations that effectively align and exploit the source content with target timbre elements.
To enable stable training and fast inference, a conditional flow-matching model is presented reconstruct the Mel-spectrogram of the source speech. 
Experimental results demonstrate that Takin-VC outperforms all baseline systems regarding naturalness, expressiveness, speaker similarity, and real-time performance. Ablation studies further validate the effectiveness of each proposed component in our framework.

\bibliographystyle{IEEEbib}
\bibliography{main}

\begin{thebibliography}{10}

\bibitem{gan2022iqdubbing}
Wendong Gan, Bolong Wen, Ying Yan, Haitao Chen, Zhichao Wang, Hongqiang Du, Lei Xie, Kaixuan Guo, and Hai Li,
\newblock ``Iqdubbing: Prosody modeling based on discrete self-supervised speech representation for expressive voice conversion,''
\newblock {\em arXiv preprint arXiv:2201.00269}, 2022.

\bibitem{tomashenko2022voiceprivacy}
Natalia Tomashenko, Xin Wang, Emmanuel Vincent, Jose Patino, Brij Mohan~Lal Srivastava, Paul-Gauthier No{\'e}, Andreas Nautsch, Nicholas Evans, Junichi Yamagishi, Benjamin O’Brien, et~al.,
\newblock ``The voiceprivacy 2020 challenge: Results and findings,''
\newblock {\em Computer Speech \& Language}, vol. 74, pp. 101362, 2022.

\bibitem{liu2021any}
Songxiang Liu, Yuewen Cao, Disong Wang, Xixin Wu, Xunying Liu, and Helen Meng,
\newblock ``Any-to-many voice conversion with location-relative sequence-to-sequence modeling,''
\newblock {\em IEEE/ACM Transactions on Audio, Speech, and Language Processing}, vol. 29, pp. 1717--1728, 2021.

\bibitem{li2023dvqvc}
Dayong Li, Xian Li, and Xiaofei Li,
\newblock ``Dvqvc: An unsupervised zero-shot voice conversion framework,''
\newblock in {\em ICASSP 2023-2023 IEEE International Conference on Acoustics, Speech and Signal Processing (ICASSP)}. IEEE, 2023, pp. 1--5.

\bibitem{hussain2023ace}
Shehzeen Hussain, Paarth Neekhara, Jocelyn Huang, Jason Li, and Boris Ginsburg,
\newblock ``Ace-vc: Adaptive and controllable voice conversion using explicitly disentangled self-supervised speech representations,''
\newblock in {\em ICASSP 2023-2023 IEEE International Conference on Acoustics, Speech and Signal Processing (ICASSP)}, 2023, pp. 1--5.

\bibitem{choi2023diff}
Ha-Yeong Choi, Sang-Hoon Lee, and Seong-Whan Lee,
\newblock ``Diff-hiervc: Diffusion-based hierarchical voice conversion with robust pitch generation and masked prior for zero-shot speaker adaptation,''
\newblock {\em International Speech Communication Association}, pp. 2283--2287, 2023.

\bibitem{anastassiou2024voiceshop}
Philip Anastassiou, Zhenyu Tang, Kainan Peng, Dongya Jia, Jiaxin Li, Ming Tu, Yuping Wang, Yuxuan Wang, and Mingbo Ma,
\newblock ``Voiceshop: A unified speech-to-speech framework for identity-preserving zero-shot voice editing,''
\newblock {\em arXiv preprint arXiv:2404.06674}, 2024.

\bibitem{luo2024posterior}
Yin-Jyun Luo and Simon Dixon,
\newblock ``Posterior variance-parameterised gaussian dropout: Improving disentangled sequential autoencoders for zero-shot voice conversion,''
\newblock in {\em ICASSP 2024-2024 IEEE International Conference on Acoustics, Speech and Signal Processing (ICASSP)}. IEEE, 2024, pp. 11676--11680.

\bibitem{wu2020vqvc}
Da-Yi Wu, Yen-Hao Chen, and Hung-Yi Lee,
\newblock ``Vqvc+: One-shot voice conversion by vector quantization and u-net architecture,''
\newblock {\em arXiv preprint arXiv:2006.04154}, 2020.

\bibitem{wu2020one}
Da-Yi Wu and Hung-yi Lee,
\newblock ``One-shot voice conversion by vector quantization,''
\newblock in {\em ICASSP 2020-2020 IEEE International Conference on Acoustics, Speech and Signal Processing (ICASSP)}. IEEE, 2020, pp. 7734--7738.

\bibitem{tang2022avqvc}
Huaizhen Tang, Xulong Zhang, Jianzong Wang, Ning Cheng, and Jing Xiao,
\newblock ``Avqvc: One-shot voice conversion by vector quantization with applying contrastive learning,''
\newblock in {\em ICASSP 2022-2022 IEEE International Conference on Acoustics, Speech and Signal Processing (ICASSP)}. IEEE, 2022, pp. 4613--4617.

\bibitem{wang2021vqmivc}
Disong Wang, Liqun Deng, Yu~Ting Yeung, Xiao Chen, Xunying Liu, and Helen Meng,
\newblock ``Vqmivc: Vector quantization and mutual information-based unsupervised speech representation disentanglement for one-shot voice conversion,''
\newblock {\em arXiv preprint arXiv:2106.10132}, 2021.

\bibitem{yang2022speech}
SiCheng Yang, Methawee Tantrawenith, Haolin Zhuang, Zhiyong Wu, Aolan Sun, Jianzong Wang, Ning Cheng, Huaizhen Tang, Xintao Zhao, Jie Wang, et~al.,
\newblock ``Speech representation disentanglement with adversarial mutual information learning for one-shot voice conversion,''
\newblock {\em arXiv preprint arXiv:2208.08757}, 2022.

\bibitem{huang2023limi}
Liangjie Huang, Tian Yuan, Yunming Liang, Zeyu Chen, Can Wen, Yanlu Xie, Jinsong Zhang, and Dengfeng Ke,
\newblock ``Limi-vc: A light weight voice conversion model with mutual information disentanglement,''
\newblock in {\em ICASSP 2023-2023 IEEE International Conference on Acoustics, Speech and Signal Processing (ICASSP)}. IEEE, 2023, pp. 1--5.

\bibitem{zhao2022disentangling}
Xintao Zhao, Feng Liu, Changhe Song, Zhiyong Wu, Shiyin Kang, Deyi Tuo, and Helen Meng,
\newblock ``Disentangling content and fine-grained prosody information via hybrid asr bottleneck features for voice conversion,''
\newblock in {\em ICASSP 2022-2022 IEEE International Conference on Acoustics, Speech and Signal Processing (ICASSP)}. IEEE, 2022, pp. 7022--7026.

\bibitem{dang2022training}
Trung Dang, Dung Tran, Peter Chin, and Kazuhito Koishida,
\newblock ``Training robust zero-shot voice conversion models with self-supervised features,''
\newblock in {\em ICASSP 2022-2022 IEEE International Conference on Acoustics, Speech and Signal Processing (ICASSP)}. IEEE, 2022, pp. 6557--6561.

\bibitem{yao2024promptvc}
Jixun Yao, Yuguang Yang, Yi~Lei, Ziqian Ning, Yanni Hu, Yu~Pan, Jingjing Yin, Hongbin Zhou, Heng Lu, and Lei Xie,
\newblock ``Promptvc: Flexible stylistic voice conversion in latent space driven by natural language prompts,''
\newblock in {\em ICASSP 2024-2024 IEEE International Conference on Acoustics, Speech and Signal Processing (ICASSP)}. IEEE, 2024, pp. 10571--10575.

\bibitem{pan2023msac}
Yu~Pan, Yuguang Yang, Yuheng Huang, Jixun Yao, Jingjing Yin, Yanni Hu, Heng Lu, Lei Ma, and Jianjun Zhao,
\newblock ``Msac: Multiple speech attribute control method for reliable speech emotion recognition,''
\newblock {\em arXiv preprint arXiv:2308.04025}, 2023.

\bibitem{pan2024gemo}
Yu~Pan, Yanni Hu, Yuguang Yang, Wen Fei, Jixun Yao, Heng Lu, Lei Ma, and Jianjun Zhao,
\newblock ``Gemo-clap: Gender-attribute-enhanced contrastive language-audio pretraining for accurate speech emotion recognition,''
\newblock in {\em ICASSP 2024-2024 IEEE International Conference on Acoustics, Speech and Signal Processing (ICASSP)}. IEEE, 2024, pp. 10021--10025.

\bibitem{pan2024gmp}
Yu~Pan, Yuguang Yang, Heng Lu, Lei Ma, and Jianjun Zhao,
\newblock ``Gmp-atl: Gender-augmented multi-scale pseudo-label enhanced adaptive transfer learning for speech emotion recognition via hubert,''
\newblock {\em arXiv preprint arXiv:2405.02151}, 2024.

\bibitem{yao2024musa}
Jixun Yao, Qing Wang, Pengcheng Guo, Ziqian Ning, Yuguang Yang, Yu~Pan, and Lei Xie,
\newblock ``Musa: Multi-lingual speaker anonymization via serial disentanglement,''
\newblock {\em arXiv preprint arXiv:2407.11629}, 2024.

\bibitem{jiang2024mega}
Ziyue Jiang, Jinglin Liu, Yi~Ren, Jinzheng He, Zhenhui Ye, Shengpeng Ji, Qian Yang, Chen Zhang, Pengfei Wei, Chunfeng Wang, et~al.,
\newblock ``Mega-tts 2: Boosting prompting mechanisms for zero-shot speech synthesis,''
\newblock in {\em The Twelfth International Conference on Learning Representations}, 2024.

\bibitem{pan2024ctefm}
Yu~Pan, Yuguang Yang, Jixun Yao, Jianhao Ye, Hongbin Zhou, Lei Ma, and Jianjun Zhao,
\newblock ``Ctefm-vc: Zero-shot voice conversion based on content-aware timbre ensemble modeling and flow matching,''
\newblock {\em arXiv preprint arXiv:2411.02026}, 2024.

\bibitem{wang2023lm}
Zhichao Wang, Yuanzhe Chen, Lei Xie, Qiao Tian, and Yuping Wang,
\newblock ``Lm-vc: Zero-shot voice conversion via speech generation based on language models,''
\newblock {\em IEEE Signal Processing Letters}, 2023.

\bibitem{borsos2023audiolm}
Zal{\'a}n Borsos, Rapha{\"e}l Marinier, Damien Vincent, Eugene Kharitonov, Olivier Pietquin, Matt Sharifi, Dominik Roblek, Olivier Teboul, David Grangier, Marco Tagliasacchi, et~al.,
\newblock ``Audiolm: a language modeling approach to audio generation,''
\newblock {\em IEEE/ACM Transactions on Audio, Speech, and Language Processing}, 2023.

\bibitem{yang2023hybridformer}
Yuguang Yang, Yu~Pan, Jingjing Yin, Jiangyu Han, Lei Ma, and Heng Lu,
\newblock ``Hybridformer: Improving squeezeformer with hybrid attention and nsr mechanism,''
\newblock in {\em ICASSP 2023-2023 IEEE International Conference on Acoustics, Speech and Signal Processing (ICASSP)}. IEEE, 2023, pp. 1--5.

\bibitem{chen2022wavlm}
Sanyuan Chen, Chengyi Wang, Zhengyang Chen, Yu~Wu, Shujie Liu, Zhuo Chen, Jinyu Li, Naoyuki Kanda, Takuya Yoshioka, Xiong Xiao, et~al.,
\newblock ``Wavlm: Large-scale self-supervised pre-training for full stack speech processing,''
\newblock {\em IEEE Journal of Selected Topics in Signal Processing}, vol. 16, no. 6, pp. 1505--1518, 2022.

\bibitem{lee2022bigvgan}
Sang-gil Lee, Wei Ping, Boris Ginsburg, Bryan Catanzaro, and Sungroh Yoon,
\newblock ``Bigvgan: A universal neural vocoder with large-scale training,''
\newblock {\em arXiv preprint arXiv:2206.04658}, 2022.

\bibitem{zen2019libritts}
Heiga Zen, Viet Dang, Rob Clark, Yu~Zhang, Ron~J Weiss, Ye~Jia, Zhifeng Chen, and Yonghui Wu,
\newblock ``Libritts: A corpus derived from librispeech for text-to-speech,''
\newblock {\em arXiv preprint arXiv:1904.02882}, 2019.

\bibitem{hsu2021hubert}
Wei-Ning Hsu, Benjamin Bolte, Yao-Hung~Hubert Tsai, Kushal Lakhotia, Ruslan Salakhutdinov, and Abdelrahman Mohamed,
\newblock ``Hubert: Self-supervised speech representation learning by masked prediction of hidden units,''
\newblock {\em IEEE/ACM transactions on audio, speech, and language processing}, vol. 29, pp. 3451--3460, 2021.

\bibitem{baevski2020wav2vec}
Alexei Baevski, Yuhao Zhou, Abdelrahman Mohamed, and Michael Auli,
\newblock ``wav2vec 2.0: A framework for self-supervised learning of speech representations,''
\newblock {\em Advances in neural information processing systems}, vol. 33, pp. 12449--12460, 2020.

\bibitem{ho2020denoising}
Jonathan Ho, Ajay Jain, and Pieter Abbeel,
\newblock ``Denoising diffusion probabilistic models,''
\newblock {\em Advances in neural information processing systems}, vol. 33, pp. 6840--6851, 2020.

\bibitem{lu2022dpm}
Cheng Lu, Yuhao Zhou, Fan Bao, Jianfei Chen, Chongxuan Li, and Jun Zhu,
\newblock ``Dpm-solver: A fast ode solver for diffusion probabilistic model sampling in around 10 steps,''
\newblock {\em Advances in Neural Information Processing Systems}, vol. 35, pp. 5775--5787, 2022.

\bibitem{li2024sef}
Junjie Li, Yiwei Guo, Xie Chen, and Kai Yu,
\newblock ``Sef-vc: Speaker embedding free zero-shot voice conversion with cross attention,''
\newblock in {\em ICASSP 2024-2024 IEEE International Conference on Acoustics, Speech and Signal Processing (ICASSP)}. IEEE, 2024, pp. 12296--12300.

\bibitem{babu2021xls}
Arun Babu, Changhan Wang, Andros Tjandra, Kushal Lakhotia, Qiantong Xu, Naman Goyal, Kritika Singh, Patrick Von~Platen, Yatharth Saraf, Juan Pino, et~al.,
\newblock ``Xls-r: Self-supervised cross-lingual speech representation learning at scale,''
\newblock {\em arXiv preprint arXiv:2111.09296}, 2021.

\bibitem{popov2021diffvc}
Vadim Popov, Ivan Vovk, Vladimir Gogoryan, Tasnima Sadekova, Mikhail Kudinov, and Jiansheng Wei,
\newblock ``Diffusion-based voice conversion with fast maximum likelihood sampling scheme,''
\newblock {\em arXiv preprint arXiv:2109.13821}, 2021.

\bibitem{choi2024dddm}
Ha-Yeong Choi, Sang-Hoon Lee, and Seong-Whan Lee,
\newblock ``Dddm-vc: Decoupled denoising diffusion models with disentangled representation and prior mixup for verified robust voice conversion,''
\newblock in {\em Proceedings of the AAAI Conference on Artificial Intelligence}, 2024, vol.~38, pp. 17862--17870.

\bibitem{zhang2023speak}
Ziqiang Zhang, Long Zhou, Chengyi Wang, Sanyuan Chen, Yu~Wu, Shujie Liu, Zhuo Chen, Yanqing Liu, Huaming Wang, Jinyu Li, et~al.,
\newblock ``Speak foreign languages with your own voice: Cross-lingual neural codec language modeling,''
\newblock {\em arXiv preprint arXiv:2303.03926}, 2023.

\bibitem{baadeneural}
Alan Baade, Puyuan Peng, and David Harwath,
\newblock ``Neural codec language models for disentangled and textless voice conversion,''
\newblock 2024.

\bibitem{defossez2022high}
Alexandre D{\'e}fossez, Jade Copet, Gabriel Synnaeve, and Yossi Adi,
\newblock ``High fidelity neural audio compression,''
\newblock {\em arXiv preprint arXiv:2210.13438}, 2022.

\bibitem{yang2023hifi}
Dongchao Yang, Songxiang Liu, Rongjie Huang, Jinchuan Tian, Chao Weng, and Yuexian Zou,
\newblock ``Hifi-codec: Group-residual vector quantization for high fidelity audio codec,''
\newblock {\em arXiv preprint arXiv:2305.02765}, 2023.

\bibitem{pan2024promptcodec}
Yu~Pan, Lei Ma, and Jianjun Zhao,
\newblock ``Promptcodec: High-fidelity neural speech codec using disentangled representation learning based adaptive feature-aware prompt encoders,''
\newblock {\em arXiv preprint arXiv:2404.02702}, 2024.

\bibitem{lipman2022flow}
Yaron Lipman, Ricky~TQ Chen, Heli Ben-Hamu, Maximilian Nickel, and Matt Le,
\newblock ``Flow matching for generative modeling,''
\newblock {\em arXiv preprint arXiv:2210.02747}, 2022.

\bibitem{tong2023improving}
Alexander Tong, Nikolay Malkin, Guillaume Huguet, Yanlei Zhang, Jarrid Rector-Brooks, Kilian Fatras, Guy Wolf, and Yoshua Bengio,
\newblock ``Improving and generalizing flow-based generative models with minibatch optimal transport,''
\newblock {\em arXiv preprint arXiv:2302.00482}, 2023.

\bibitem{tong2023simulation}
Alexander Tong, Nikolay Malkin, Kilian Fatras, Lazar Atanackovic, Yanlei Zhang, Guillaume Huguet, Guy Wolf, and Yoshua Bengio,
\newblock ``Simulation-free schr$\backslash$" odinger bridges via score and flow matching,''
\newblock {\em arXiv preprint arXiv:2307.03672}, 2023.

\bibitem{bartosh2023neural}
Grigory Bartosh, Dmitry Vetrov, and Christian~A Naesseth,
\newblock ``Neural diffusion models,''
\newblock {\em arXiv preprint arXiv:2310.08337}, 2023.

\bibitem{zhou2023denoising}
Linqi Zhou, Aaron Lou, Samar Khanna, and Stefano Ermon,
\newblock ``Denoising diffusion bridge models,''
\newblock {\em arXiv preprint arXiv:2309.16948}, 2023.

\bibitem{liu2023generative}
Alexander~H Liu, Matt Le, Apoorv Vyas, Bowen Shi, Andros Tjandra, and Wei-Ning Hsu,
\newblock ``Generative pre-training for speech with flow matching,''
\newblock {\em arXiv preprint arXiv:2310.16338}, 2023.

\bibitem{kim2024p}
Sungwon Kim, Kevin Shih, Joao~Felipe Santos, Evelina Bakhturina, Mikyas Desta, Rafael Valle, Sungroh Yoon, Bryan Catanzaro, et~al.,
\newblock ``P-flow: a fast and data-efficient zero-shot tts through speech prompting,''
\newblock {\em Advances in Neural Information Processing Systems}, vol. 36, 2024.

\bibitem{ronneberger2015u}
Olaf Ronneberger, Philipp Fischer, and Thomas Brox,
\newblock ``U-net: Convolutional networks for biomedical image segmentation,''
\newblock in {\em Medical image computing and computer-assisted intervention--MICCAI 2015: 18th international conference, Munich, Germany, October 5-9, 2015, proceedings, part III 18}. Springer, 2015, pp. 234--241.

\bibitem{gulati2020conformer}
Anmol Gulati, James Qin, Chung-Cheng Chiu, Niki Parmar, Yu~Zhang, Jiahui Yu, Wei Han, Shibo Wang, Zhengdong Zhang, Yonghui Wu, et~al.,
\newblock ``Conformer: Convolution-augmented transformer for speech recognition,''
\newblock {\em arXiv preprint arXiv:2005.08100}, 2020.

\bibitem{yang2022lmec}
Yuguang Yang, Yu~Pan, Jingjing Yin, and Heng Lu,
\newblock ``Lmec: Learnable multiplicative absolute position embedding based conformer for speech recognition,''
\newblock {\em arXiv preprint arXiv:2212.02099}, 2022.

\bibitem{kim2022squeezeformer}
Sehoon Kim, Amir Gholami, Albert Shaw, Nicholas Lee, Karttikeya Mangalam, Jitendra Malik, Michael~W Mahoney, and Kurt Keutzer,
\newblock ``Squeezeformer: An efficient transformer for automatic speech recognition,''
\newblock {\em Advances in Neural Information Processing Systems}, vol. 35, pp. 9361--9373, 2022.

\bibitem{perez2018film}
Ethan Perez, Florian Strub, Harm De~Vries, Vincent Dumoulin, and Aaron Courville,
\newblock ``Film: Visual reasoning with a general conditioning layer,''
\newblock in {\em Proceedings of the AAAI conference on artificial intelligence}, 2018, vol.~32.

\bibitem{lin2021s2vc}
Jheng-hao Lin, Yist~Y Lin, Chung-Ming Chien, and Hung-yi Lee,
\newblock ``S2vc: A framework for any-to-any voice conversion with self-supervised pretrained representations,''
\newblock {\em arXiv preprint arXiv:2104.02901}, 2021.

\bibitem{tong2023conditional}
Alexander Tong, Nikolay Malkin, Guillaume Huguet, Yanlei Zhang, Jarrid Rector-Brooks, Kilian Fatras, Guy Wolf, and Yoshua Bengio,
\newblock ``Conditional flow matching: Simulation-free dynamic optimal transport,''
\newblock {\em arXiv preprint arXiv:2302.00482}, vol. 2, no. 3, 2023.

\bibitem{shen2023naturalspeech2}
Kai Shen, Zeqian Ju, Xu~Tan, Yanqing Liu, Yichong Leng, Lei He, Tao Qin, Sheng Zhao, and Jiang Bian,
\newblock ``Naturalspeech 2: Latent diffusion models are natural and zero-shot speech and singing synthesizers,''
\newblock {\em arXiv preprint arXiv:2304.09116}, 2023.

\bibitem{wang2023valle}
Chengyi Wang, Sanyuan Chen, Yu~Wu, Ziqiang Zhang, Long Zhou, Shujie Liu, Zhuo Chen, Yanqing Liu, Huaming Wang, Jinyu Li, et~al.,
\newblock ``Neural codec language models are zero-shot text to speech synthesizers,''
\newblock {\em arXiv preprint arXiv:2301.02111}, 2023.

\bibitem{yao2024stablevc}
Jixun Yao, Yuguang Yan, Yu~Pan, Ziqian Ning, Jiaohao Ye, Hongbin Zhou, and Lei Xie,
\newblock ``Stablevc: Style controllable zero-shot voice conversion with conditional flow matching,''
\newblock {\em arXiv preprint arXiv:2412.04724}, 2024.

\bibitem{liu2024seedvc}
Songting Liu,
\newblock ``Zero-shot voice conversion with diffusion transformers,''
\newblock {\em arXiv preprint arXiv:2411.09943}, 2024.

\bibitem{he2016identity}
Kaiming He, Xiangyu Zhang, Shaoqing Ren, and Jian Sun,
\newblock ``Identity mappings in deep residual networks,''
\newblock in {\em Computer Vision--ECCV 2016: 14th European Conference, Amsterdam, The Netherlands, October 11--14, 2016, Proceedings, Part IV 14}. Springer, 2016, pp. 630--645.

\bibitem{van2008visualizing}
Laurens Van~der Maaten and Geoffrey Hinton,
\newblock ``Visualizing data using t-sne.,''
\newblock {\em Journal of machine learning research}, vol. 9, no. 11, 2008.

\end{thebibliography}

\end{document}